\documentclass[english,aps,preprint]{revtex4}
\usepackage[LGR,T1]{fontenc}
\usepackage[latin9]{inputenc}
\usepackage{float}
\usepackage{amsmath}
\usepackage{graphicx}

\makeatletter

\DeclareRobustCommand{\greektext}{%
  \fontencoding{LGR}\selectfont\def\encodingdefault{LGR}}
\DeclareRobustCommand{\textgreek}[1]{\leavevmode{\greektext #1}}
\ProvideTextCommand{\~}{LGR}[1]{\char126#1}

\@ifundefined{textcolor}{}
{%
 \definecolor{BLACK}{gray}{0}
 \definecolor{WHITE}{gray}{1}
 \definecolor{RED}{rgb}{1,0,0}
 \definecolor{GREEN}{rgb}{0,1,0}
 \definecolor{BLUE}{rgb}{0,0,1}
 \definecolor{CYAN}{cmyk}{1,0,0,0}
 \definecolor{MAGENTA}{cmyk}{0,1,0,0}
 \definecolor{YELLOW}{cmyk}{0,0,1,0}
}

\makeatother

\usepackage{babel}
\begin{document}
\title{The Phase Transition of Reissner-Nordstr$\ddot{o}$m Black Holes}
\author{Bobin Li}
\email{gslibobin@lut.edu.cn}

\affiliation{Department of physics, School of Science, Lanzhou University of Technology,
Lanzhou 730050, P.R.China }
\begin{abstract}
Under the framework of thermodynamics, the phase transition
of the black hole is a general issue in general relativity. In this
work, the phase transition of charged black holes is discussed carefully.
The metric tensor of thermodynamics is redefined in the charged black
hole, based on the Ruppeiner geometry. With the well-defined metric
tensor of thermodynamics, the scalar curvature of the charged black
hole is obtained. It is indicated that the scalar curvature is diverged
and infinite when the mass M or charge Q are set by some values, and
it is shown that the charged black hole suffers from a phase transition.
At the same time, there is a phase transition from small mass to large
mass or from small to high charged state. It is shown that the phase
transition of a charged black hole is a common and general process
and this work is meaningful for the construction of microscopic states
of black holes.
\end{abstract}
\maketitle

\section{introduction}

Under the framework of thermodynamics, the phase transition of the
black hole is a general issue in general relativity. The thermodynamics
and stability of charged black holes in anti-de Sitter space have
been extensively studied in the literature. Fernando \cite{1} focused
on the thermodynamics of static electrically charged Born-Infeld black
holes in AdS space, noting that small black holes are unstable while
large black holes are stable. Myung et al.  \cite{2} compared the thermodynamic
quantities of Born-Infeld-anti-de Sitter black holes with those of
Reissner-Norstr$\ddot{o}$m-AdS and Schwarzschild-AdS black holes.
Niu et al.  \cite{3} studied the phase transition of Reissner-Nordstr$\ddot{o}$m
black holes in anti-de Sitter spacetime, finding identical exponents
for an RN-AdS black hole and a Van der Waals liquid gas system. Wei
et al.  \cite{4,7} explored the phase structure and equilibrium state
space geometry of charged topological Gauss-Bonnet black holes in
anti-de Sitter spacetime. Cai et al.  \cite{5} investigated the P-V
criticality and phase transition in the extended phase space of charged
Gauss-Bonnet black holes in AdS space, highlighting the appearance
of criticality and phase transition in certain dimensions and conditions.
Ma et al.  \cite{6} discussed the existence condition and second-order
phase transition of RN-dS black holes. Hendi et al.  \cite{9} studied
the phase transition and thermodynamic geometry of Einstein-Maxwell-dilaton
black holes, identifying two types of phase transitions based on parameter
values. Xu et al. \cite{8} examined the P-V criticality and phase transition
in the extended phase space of black holes in massive gravity, showing
a black hole-to-vacuum phase transition in certain cases. Overall,
these studies contribute to a deeper understanding of the phase transition
behavior of charged black holes in anti-de Sitter spacetime  \cite{10}. 

The phase transition of black holes has been a topic of interest
in various gravitational theories. Myung  \cite{11} studied the phase
transition between black holes with scalar hair and topological black
holes, showing that the transition is second-order based on differences
in free energies. Wei et al.  \cite{12} explored the thermodynamic geometry
and phase transition of the Kehagias-Sfetsos black hole in deformed
Ho\v{r}ava-Lifshitz gravity, suggesting an effective repulsive interaction
similar to an ideal gas of fermions. Ma  \cite{13} investigated the
thermodynamic phase transition of black holes in asymptotically safe
gravity by introducing spatially finite boundary conditions and found
similarities to the thermodynamic phase structure of RN-AdS black
holes. Galante et al.  \cite{14} discussed a phase transition driven
by chemical potential in Einstein-Maxwell theory conformally coupled
to a scalar field, leading to an asymptotically anti-de Sitter black
hole. Hennigar et al.  \cite{15} presented a phase transition in asymptotically
anti-de Sitter hairy black holes in Lovelock gravity, showcasing a
\textquotedbl\textgreek{l}-line\textquotedbl{} phase transition
in black hole thermodynamics. Feng et al.  \cite{16} studied the thermodynamics
and phase transition of a rainbow Schwarzschild black hole in rainbow
gravity, analyzing critical behavior and thermodynamic stability.
Rodrigue et al.  \cite{17} investigated the thermodynamics and Hawking
radiation of Schwarzschild black holes with quintessence-like matter
and deficit solid angle, deriving expressions for temperature and
specific heat. Bhattacharya et al.  \cite{18} focused on the extremal
phase transition of black holes, showing that the phase transition
occurs only in the microcanonical ensemble and calculating critical
exponents. Xu et al.  \cite{19} extended the study of the relationship
between the photon sphere and thermodynamic phase transition to charged
Born-Infeld-AdS black holes, identifying different cases of phase
transitions based on the Born-Infeld parameter. Lastly, Cao et al.
 \cite{20} discussed the thermodynamic phase structure of the complex
Sachdev-Ye-Kitaev model and charged black holes in deformed JT gravity,
drawing parallels to the van der Waals-Maxwell phase transition theory.
Overall, these studies contribute to our understanding of the phase
transitions of black holes in various gravitational theories.

Chaloshtary et al.  \cite{21} explored the critical behavior and reentrant
phase transition for logarithmic nonlinear charged black holes in
massive gravity. Wu et al.  \cite{22} identified a crucial parameter
that characterizes the thermodynamic properties of black holes in
massive gravity, affecting the existence of different black hole phases.
Zhou et al.  \cite{23} investigated the impact of the generalized uncertainty
principle on black hole thermodynamics and phase transitions in a
cavity, observing the Gross-Perry-Yaffe phase transition for both
large and small black hole configurations. Furthermore, Zhou et al.
 \cite{24}also studied the phase transition and microstructures of
five-dimensional charged Gauss-Bonnet-AdS black holes in the grand
canonical ensemble, highlighting the constancy of interaction at the
phase transition. Li et al. \cite{25} proposed that the kinetics of
Hawking-Page phase transitions could be governed by a reaction-diffusion
equation, with Hawking evaporation playing a crucial role in facilitating
the phase transition. Xu et al. {\cite{26} demonstrated how the Landau
functional can reflect the physical process of black hole phase transitions,
with the splitting and transformation of the global minimum corresponding
to different types of phase transitions. Lastly, Cao et al.  investigated the 
thermodynamic phase structure of the complex Sachdev-Ye-Kitaev
model and charged black holes in deformed JT gravity, drawing parallels
between these models and the van der Waals-Maxwell liquid-gas model.
The presence of charge was found to influence the Hawking-Page phase
transition in neutral black holes. Overall, these studies contribute
to our understanding of the phase transitions of black holes in various
theoretical frameworks and gravitational backgrounds.

This work meticulously examines the phase transition of charged black
holes. The metric tensor of thermodynamics is reinterpreted in the
charged black hole using the Ruppeiner geometry. The scalar curvature
of a charged black hole can be calculated using the well-defined thermodynamic
metric tensor. When the mass M or charge Q is set to certain levels,
the scalar curvature diverges and becomes infinite, and the charged
black hole undergoes a phase transition. At the same moment, a phase
transition occurs from small mass to large mass, or from low to high
charged state. The section II and III reveal that the metric tensor
and entropy of Reissner-Nordstr$\ddot{o}$m black holes, and the section
IV introduce the ruppeiner geometry and metric tensor of thermodynamics.
Finally, in the section V, the scalar curvature of thermodynamics
is calculated, based on metric tensor of thermodynamics. It is demonstrated
that a charged black hole's phase transition is a common and general
phenomenon. and this work is meaningful for the construction of microscopic
states of black holes.

\section{The general black hole}

\begin{align}
ds^{2} & =-(1-\frac{2M}{r}+\frac{Q^{2}}{r^{2}})dt^{2}+(1-\frac{2M}{r}+\frac{Q^{2}}{r^{2}})^{-1}dr^{2}+r^{2}d\theta^{2}+r^{2}sin^{2}\theta d\varphi^{2}
\end{align}
The equation (1) is metrics tensor of Reissner-Nordstr$\ddot{o}$m
black holes where $(M,Q)$ is mass and charge of black hole

\section{The entropy of charged black hole}

The entropy of Reissner-Nordstr$\ddot{o}$m black holes as follows
\begin{equation}
S(M,Q)=\pi[(M+\sqrt{M^{2}-Q^{2}})^{2}]
\end{equation}

\section{Ruppeiner geometry}

According to the thermodynamic fluctuation theory

\begin{equation}
S=S_{o}+\frac{1}{2}(\partial_{\mu}\partial_{\nu}S)|_{0}\Delta x^{\mu}\Delta x^{\nu}+\cdots
\end{equation}
According to Ruppeiner geometry, the metric tensor of thermodynamics
is as follows

\begin{equation}
g_{\mu\nu}=\frac{\partial^{2}S}{\partial x_{\mu}\partial x_{\nu}}\;\;\;where\;x_{\mu},x_{\nu}=M,Q\;(\mu,\nu=1,2)
\end{equation}
And

\begin{equation}
[g_{\mu\nu}]=\left[\begin{array}{cc}
\frac{\partial^{2}S}{\partial M\partial M} & \frac{\partial^{2}S}{\partial M\partial Q}\\
\frac{\partial^{2}S}{\partial Q\partial M} & \frac{\partial^{2}S}{\partial Q\partial Q}
\end{array}\right]
\end{equation}

\section{Scalar curvature R}

\subsection{Riemannian Geometry}

According to basic general relativity, the scalar curvature is as
follows

\begin{align}
R & =-\frac{1}{\sqrt{g}}[\frac{\partial}{\partial M}(\frac{g_{QM}}{g_{MM}}\frac{1}{\sqrt{g}}\frac{\partial g_{MM}}{\partial Q}-\frac{1}{\sqrt{g}}\frac{\partial g_{QQ}}{\partial M})\nonumber \\
 & +\frac{\partial}{\partial Q}(\frac{2}{\sqrt{g}}\frac{\partial g_{QM}}{\partial Q}-\frac{1}{\sqrt{g}}\frac{\partial g_{MM}}{\partial Q}-\frac{g_{MQ}}{g_{MM}}\frac{1}{\sqrt{g}}\frac{\partial g_{MM}}{\partial M})]
\end{align}

\subsection{Scalar curvature of thermodynamics}

\begin{align}
R(M,Q) & =\frac{3M^{2}{\left(6M^{8}+6M^{7}Q-27M^{6}Q^{2}-30M^{5}Q^{3}-M^{4}Q^{4}\right)}}{Q{\left(M-Q\right)}^{2}{\left(M+Q\right)}\sigma_{1}}\nonumber \\
 & +\frac{3M^{2}{\left(34M^{3}Q^{5}+47M^{2}Q^{6}+6MQ^{7}+15Q^{8}\right)}}{Q{\left(M-Q\right)}^{2}{\left(M+Q\right)}\sigma_{1}}\nonumber \\
 & -\frac{6M{\left(M+Q\right)}^{2}{\left(3M^{5}-6M^{2}Q^{3}+MQ^{4}-2Q^{5}\right)}}{Q\sqrt{M^{2}-Q^{2}}\sigma_{1}}
\end{align}
where$\sigma_{1}={\left(-3M^{4}+6M^{2}Q^{2}+Q^{4}\right)}^{2}$

\section{Result and discussion}

\subsection{Phase transition from small mass to large mass}

\begin{figure}[H]
\begin{centering}
\includegraphics[scale=0.15]{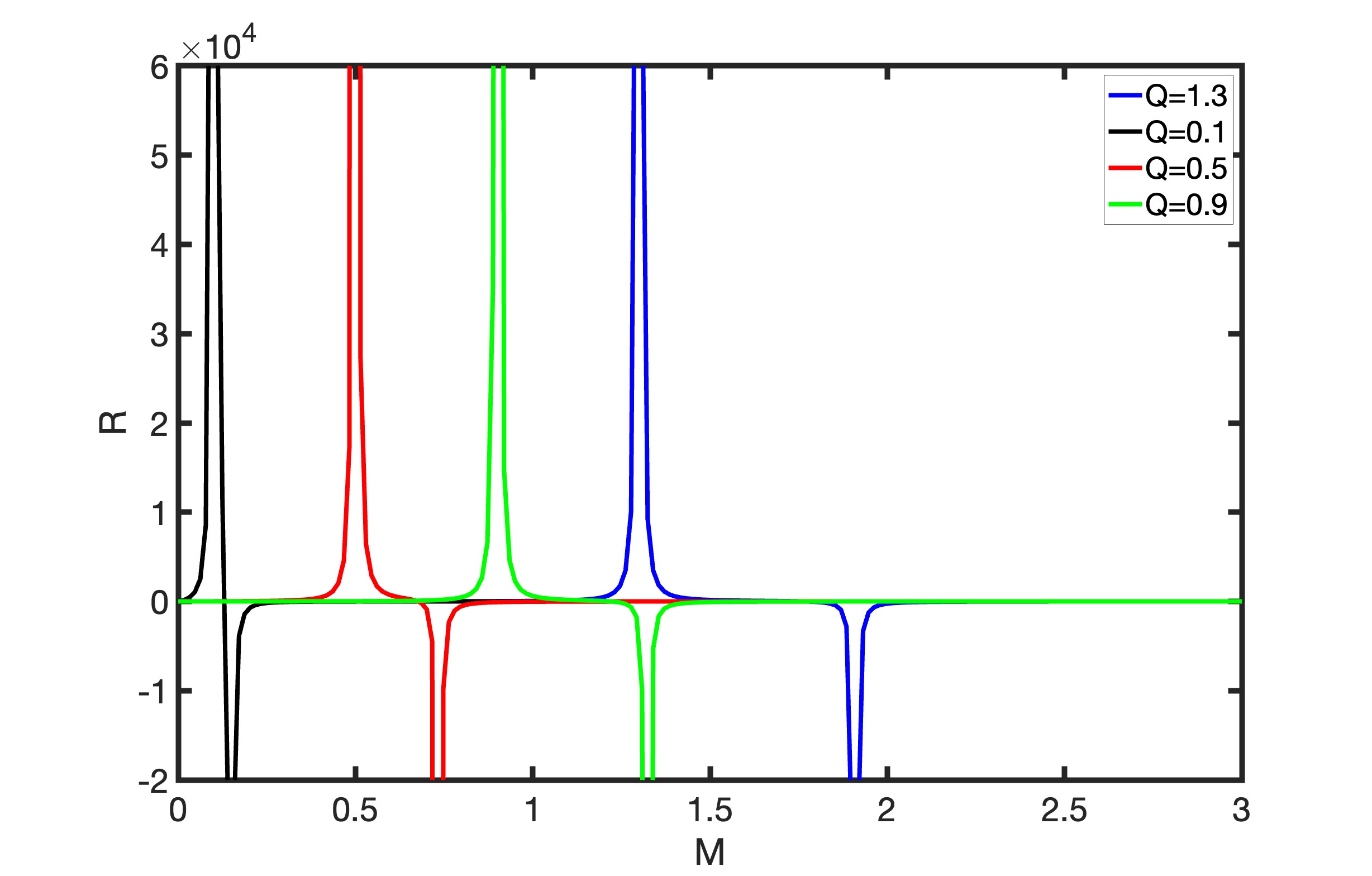}
\par\end{centering}
\caption{The diagram of scalar curvature $R$ with mass $M$ ($M>Q$){.
The critical mass}$M_{C}(Q=0.1)=0.15,$$M_{C}(Q=0.5)=0.75,$$M_{C}(Q=0.9)=1.35,$$M_{C}(Q=1.3)=1.95$}

\end{figure}

When the charge $Q\rightarrow0$, that is the case of Schwarzschild
black hole, there will be no phase transition and the critical mass
$M_{C}\rightarrow0$. For given value of charge $Q$ of black hole,
there are one critical mass $M_{C}$ corresponding to the divergence
of scalar curvature $R$. And it is indicated that there is a phase
transition from small mass to big mass for a charged black hole. 

Due to$M>Q$ of charged black hole, scalar curvature $R$ actually
has one critical point to be divergent. For different charged states
of black hole, there are different critical mass $M_{C}$ of phase
transition. Shown in Fig.1, $M_{C}(Q=0.1)=0.15,$ $M_{C}(Q=0.5)=0.75,$
$M_{C}(Q=0.9)=1.35,$ $M_{C}(Q=1.3)=1.95$.

\subsection{Phase transition from small charged to high charged state}

\begin{figure}[H]
\begin{centering}
\includegraphics[scale=0.15]{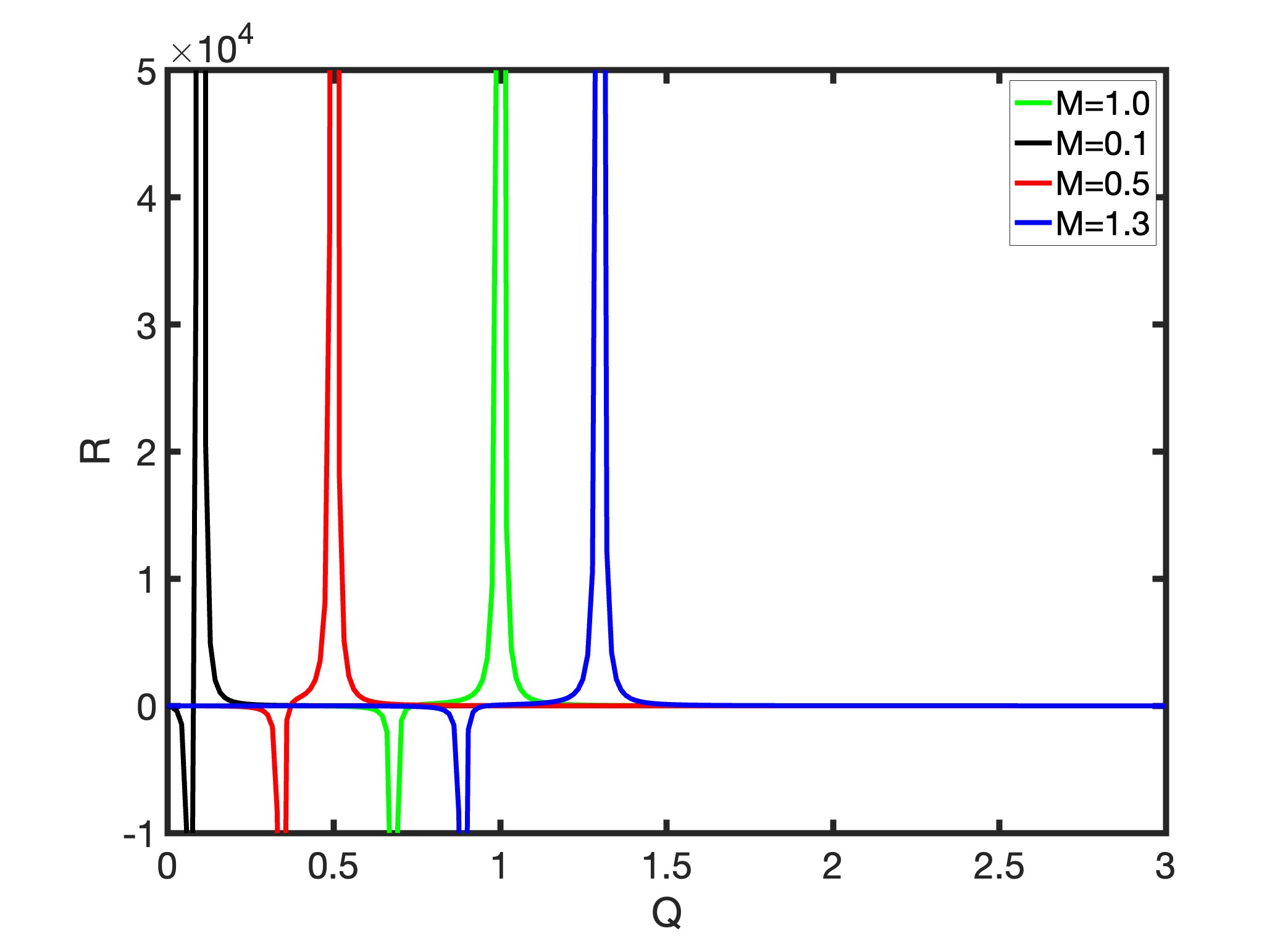}
\par\end{centering}
\caption{The diagram of scalar curvature $R$ with charged $Q(Q<M)${.
The critical charge}$Q_{C}(M=0.1)=0.07,$$Q_{C}(M=0.5)=0.33,$$Q_{C}(M=1.0)=0.67,$$Q_{C}(M=1.3)=0.87$}

\end{figure}

For given value of mass $M$ of black hole, there are one critical
charge $Q_{C}$ corresponding to the divergence of scalar curvature
$R$. And it is indicated that there is an phase transition from small
charged to high charged state for a charged black hole.

Due to $M>Q$ of charged black hole, scalar curvature $R$ actually
has one critical point to be divergent. For different mass of black
hole, there are different critical charge $Q_{C}$ of phase transition.
Shown in Fig.2, $Q_{C}(M=0.1)=0.07,$ $Q_{C}(M=0.5)=0.33,$ $Q_{C}(M=1.0)=0.67,$
$Q_{C}(M=1.3)=0.87$.

\subsection{Phase transition Overview}

\begin{figure}[H]
\begin{centering}
\includegraphics[scale=0.15]{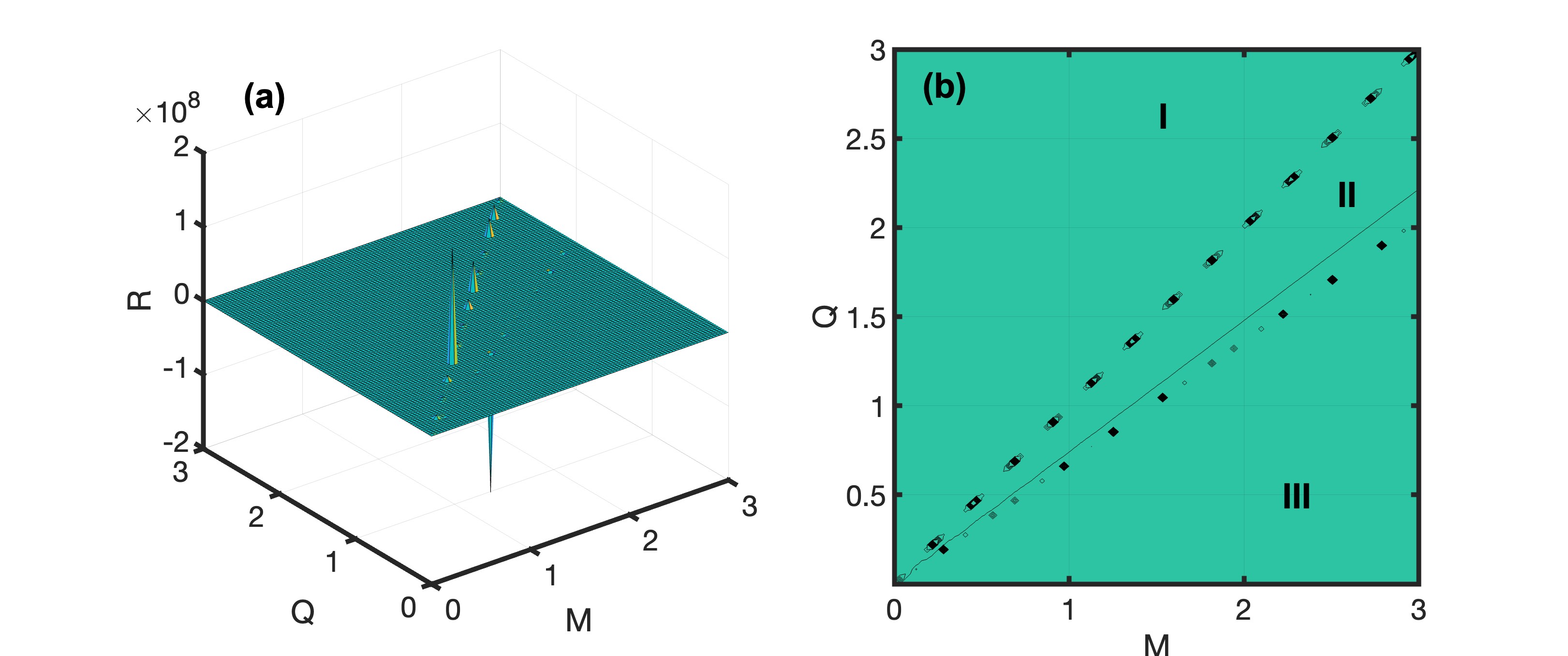}
\par\end{centering}
\caption{The diagram of scalar curvature R with mass M and charge Q}

\end{figure}

From overview diagram $R\sim QM$, there are one critical lines on
plane $QM(M>Q)$. And the critical line means that the phase of charged
black hole would be transform from one state to another state. Figure
3(b) shows that there are three phase zone I ($M>Q$,impossible zone),
zone II and zone III. For fixed charge $Q$ or fixed mass $M$ of
charged black hole, the phase state will transfer from phase zone
II to zone III. And this situation is like the phase transition of the ordinary matter, 
for example of transition between liquid and  gas phase.

The critical lines on plane $QM$ means where the scalar curvature
$R$ will diverge and infinite. But it is shown that it is discontinuous
and a series of islands for where scalar curvature $R$ is divergent,
and it is interesting features for phase transition of charged black
hole. The critical lines is 
\begin{equation}
M_{c}=\frac{3}{2}Q\;\;\;and\;\;\;Q_{c}=\frac{2}{3}M
\end{equation}

In the usual situation, the value of scalar curvature $R$ is equal
to zero, and it indicates that there is no interaction between the
microscopic elements of the charged black hole, and the microscopic
elements are free and are like the ideal gas under the condition of
thermodynamic fluctuation theory. In the critical point of phase transition,
the value of scalar curvature $R$ is less than $0$ $(R<0)$, and
it is indicated that there is an attractive force between microscopic
elements of the charged black hole. the more approach the critical
point, the stronger this attractive force. The correlation length
will increase from local to global between microscopic elements of
the charged black hole. and it is shown that the microscopic structure
of the charged black hole would rearrange and transform from one phase
to another phase.

\subsection{The critical exponents of scalar curvature $R$}

Under the condition of $M_{c}=\frac{3}{2}Q$, scalar curvature $R$
is expand as follows

\begin{align}
R(M)|_{M=\frac{3}{2}Q}\approx & const_{2}(Q)\cdot(M-\frac{3}{2}Q)^{1}-const_{3}(Q)\cdot(M-\frac{3}{2}Q)^{2}+O(2)
\end{align}
Under the condition of $M_{c}=\frac{3}{2}Q$, 

\begin{equation}
\ensuremath{R(M)|_{M=\frac{3}{2}Q}\approx const(Q)\cdot(M-\frac{3}{2}Q)^{\nu}+O(\nu)}
\end{equation}
the critical exponents of scalar curvature $R$ is $\nu=1$

\section{Conclusion}
The black hole phase transition is a general relativity problem within the domain of thermodynamics. 
The phase transition of Reissner-Nordstr$\ddot{o}$m black holes is thoroughly examined in this article. 
Based on the Ruppeiner geometry, the metric tensor of thermodynamics is redefined in the
Reissner-Nordstr$\ddot{o}$m black holes. The charged black hole's scalar curvature can be found using 
the well-defined metric tensor of thermodynamics. It is demonstrated that the charged black hole experiences
 a phase transition where the scalar curvature is divergent and infinite, under the conditions that the mass $M$
  or charge $Q$ are fixed by certain values, that there is a phase transition from a small to a large mass or from
 a small to a highly charged state of Reissner-Nordstr$\ddot{o}$m black holes .

When the charge $Q\rightarrow0$, the critical mass $M_{C}\rightarrow0$ and there won't be a phase transition.
There is one critical masses $M_{C}$ for a fixed black hole charge $Q$, which correspond to the divergence of scalar curvature $R$.
Additionally, the critical line indicates that a charged black hole's phase would change from one state to another state. 
In phase parameter space, It demonstrates that there are three phases parts: I(impossible zone), II, and III. 
The phase state of a charged black hole will go from phase zone II to zone III for fixed charge $Q$ or fixed mass $M$. 
This situation is like the phase transition of ordinary matter, such as from liquid to gas phase. 
In the area of phase zone II or zone III, the value of scalar curvature $R$ is equal to zero, which means that there is no interaction between the
microscopic elements of Reissner-Nordstr$\ddot{o}$m black holes and resemble an ideal gas under the condition of thermodynamic fluctuation theory.
 When the scalar curvature R is smaller than zero $(R<0)$ at the crucial moment of phase transition,
it suggests that the microscopic elements of Reissner-Nordstr$\ddot{o}$m black holes are attracted to each other, and this attracting 
force becomes stronger the closer one gets to the critical point.  The microscopic elements of Reissner-Nordstr$\ddot{o}$m black holes 
will see an increase in correlation length from local to global. It is shown that the microscopic structure of the charged black hole would 
rearrange and transform from one phase to another phase.

As a conclusion, It is shown that the phase transition of a charged black hole is a common and general process and this work is 
meaningful for the construction of microscopic states of black holes.

\bibliography{apssamp}

\end{document}